\documentclass[aps,prb,showpacs,showkeys, preprint]{revtex4-1}
\usepackage{amsmath}
\usepackage{amsfonts}
\usepackage[dvips]{graphicx}
\usepackage{amssymb}

\begin{document}
\title{An exactly soluble model of a shallow double well}
\author{R Mu\~{n}oz-Vega}
\email{rodrigo.munoz@uacm.edu.mx}
\author{E L\'{o}pez-Ch\'{a}vez}\email{elopezc@hotmail.com}
\affiliation{Universidad Aut\'{o}noma de la Ciudad de M\'{e}xico,
Centro Hist\'{o}rico, Fray Servando Teresa de Mier 92,
Col. Centro, Del. Cuauht\'{e}moc, M\'{e}xico DF, CP 06080}
\author{E Salinas-Hern\'{a}ndez}\email{esalinas@ipn.mx}
\affiliation{ESCOM-IPN, Av Juan de Dios B\'{a}tiz s/n, Unidad Profesional Adolfo L\'{o}pez Mateos, Col Lindavista, Del G A Madero, M\'{e}xico DF, CP 07738}
\author{J-J Flores-Godoy}\email{job.flores@ibero.mx}
\author{G Fern\'{a}ndez-Anaya}\email{guillermo.fernandez@ibero.com}
\affiliation{Departamento de F\'isica y Matem\'aticas, Universidad Iberoamericana, Prol Paseo de la Reforma 880, Col Lomas de Santa Fe, Del A Obreg\'on, M\'exico D F, CP 01219}
\begin{abstract}
Shallow one-dimensional double well potentials appear in atomic and molecular physics and other fields. Unlike the ``deep"  wells of macroscopic quantum coherent systems, shallow double wells need not present low-lying two-level systems. We argue that this feature, the absence of a low-lying two-level system in certain shallow double wells, may allow the finding of new test grounds for quantum mechanics in mesoscopic systems. We illustrate the above ideas with a family of shallow double wells  obtained from Bargman potentials through the Darboux-B\"{a}cklund transform.
\end{abstract}
\date{\today}
\maketitle
\section{Introduction}
In the present Letter we study the properties of exactly soluble model consisting of a structureless particle on the line subject to a symmetric shallow double well potential, where by ``shallow" we mean that the ground energy level is either above or slightly below the energy of the classical separatrix ($V(x=0)$.) The main purpose of the present Letter is to elucidate in which aspects the behaviour of such shallow double wells differ from that of the ``deep" double wells that appear, for instance, in the description of Josephson junction based circuits exhibiting macroscopic quantum coherence (MQC).\cite{Leggett80, Friedmanetal2000, Cleland2012} Recently, a symmetric shallow double well potential has been used in the description of unstable phonon modes of the layered superconductor LaO$_{0.5}$F$_{0.5}$BiS$_{2}$. The model strongly suggest that a dynamical deformation of the structure of the material lies at the heart of its superconductor transition.\cite{Wanetal} Previously a similar model, leading to similar conclusions, was proposed for the superconducting composite MgCNi$_{3}$. \cite{Ignatov03, Heid04} The properties of both partially fragmented Bose-Einstein Condensates\cite{Grond2013} and high-spin fermionic systems\cite{Jurgensen2013} in shallow double well optical traps are another example of current interest. Symmetric shallow double wells may also play a role in the well superionic transition of the AgI compound, \cite{Rakitin96, Kozyrev2010}  as we shall argue in the following pages.

 In the following pages we have resorted to the Darboux-B\"{a}cklund Transform\cite{Darboux} (DBT) in order to generate exactly solvable potentials. The authors  have not been able to find any precedent in the literature where the DBT has been used specifically for the generation of shallow double wells although there is at least one study of DBT-generated asymmetric double wells, which is explicitly focused on deep wells in the context of super-symmetric quantum mechanics. \cite{Gangopadhyaya93} Double and multiple well potentials have also been generated in higher order super-symmetric quantum mechanics\cite{Fernandez04} with techniques closely related to the DBT,  but seemingly without paying no particular attention to shallow wells.

The rest of this Letter is structured as follows: In Section 2 we describe the procedure we have used for constructing soluble shallow double wells, and describe some of the features of the resulting potentials. The significance of our results is discussed in Section 2 and finally, some tentative conclusions are advanced in Section 3.   
\section{Procedure and results}

Exactly soluble double well potentials can be generated by a quasi-isospectral DBT starting from the dimensionless Hamiltonian 
\begin{equation}\label{DBT01}
\eta=-\frac{d^{2}}{dx^{2}}-2\textrm{sech}^{2} x
\end{equation}
by taking $\epsilon<0$, the factorization energy  of the transform,  below the ground level, $E_{0}^{\prime}=-1$, of Hamiltonian (\ref{DBT01}). 
We start by factorizing\cite{Scho, InfeldHull} operator $\eta$ in the form
\begin{equation}\label{DBT02}
\eta=A_{\epsilon}^{\dagger}A_{\epsilon}+\epsilon,
\end{equation}
with the use of the linear first order operators
\begin{equation}\label{DBT03}
A_{\epsilon}=-\frac{d}{dx}+\bigg(\frac{u_{\epsilon}^{\prime}}{u_{\epsilon}}\bigg)(x)
 \end{equation}
and
\begin{equation}\label{DBT04}
A_{\epsilon}^{\dagger}=\frac{d}{dx}+\bigg(\frac{u_{\epsilon}^{\prime}}{u_{\epsilon}}\bigg)(x).
\end{equation}
In (\ref{DBT03}) and (\ref{DBT04}) the prime stands for the $x$ derivative, and the so called \emph{seed function }$u_{\epsilon}:\mathbb{R}\rightarrow\mathbb{R}$, given by 
\begin{equation}\label{DBT04}
u_{\epsilon}(x)=\sinh (\sqrt{\vert\epsilon\vert}\ x)\tanh (x) -\sqrt{\vert\epsilon\vert}\cosh (\sqrt{\vert\epsilon\vert}\  x),
\end{equation}
 is a non-normalizable (thus unphysical) yet node-free solution of the eigenvalue equation
\begin{equation}
\eta u_{\epsilon}(x)=\epsilon u_{\epsilon}(x).
\end{equation}  

The new Hamiltonian 
\begin{equation}\label{DBT05}
\Xi_{\epsilon}=A_{\epsilon}A_{\epsilon}^{\dagger}+\epsilon ,
\end{equation}
which is of the usual Schroedinger form:
\begin{equation}\label{DBT05A}
\Xi_{\epsilon}=-\frac{d^{2}}{dx^{2}}+V_{\epsilon}(x)
\end{equation}
with a potential $V_{\epsilon}:\mathbb{R}\rightarrow\mathbb{R}$ that can be readily expressed in terms  of $u_{\epsilon}$ as
\begin{equation}\label{DBTV}
V_{\epsilon}=2\Bigg(\frac{u_{\epsilon}^{\prime}}{u_{\epsilon}}\Bigg)^{2}-\frac{u_{\epsilon}^{\prime\prime}}{u_{\epsilon}}+\epsilon,
\end{equation} 
has two bounded states, namely, the ground state
\begin{equation}\label{GROUND}
\psi_{0}^{(\epsilon)}(x)=\frac{1}{u_{\epsilon}(x)}\times\Bigg(\int_{-\infty}^{\infty}\frac{dx}{u_{\epsilon}^{2}(x)}\ \Bigg)^{-1/2}
\end{equation}
\begin{figure}[!htb]
\centering
\includegraphics[scale=1.2]{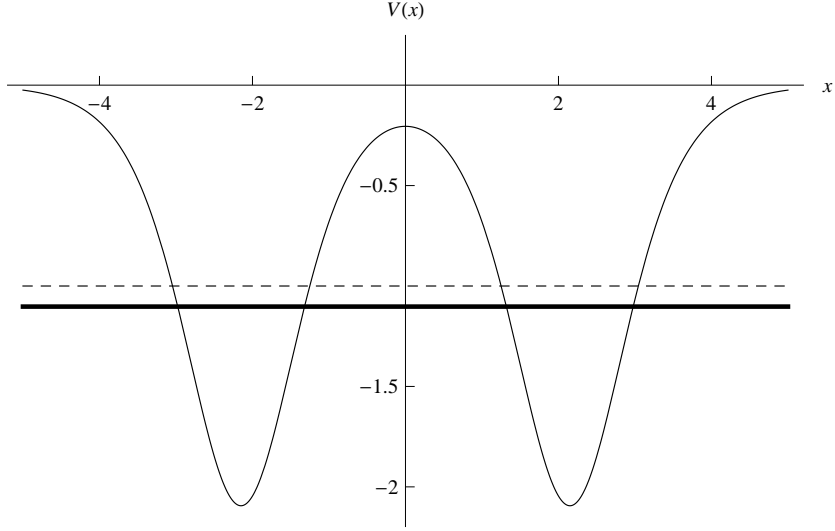}
\caption{An example of a $V_{k,\epsilon}$ potential where the ground level lies below the central barrier. In this case $-k^{2}=-1$ (dashed line) and $\epsilon=-1.10$ (thick line.)}
\label{fig:02NTM}
\end{figure}
with corresponding energy eigenvalue $E_{0}=\epsilon$, \emph{i. e.}
\begin{equation}\label{DBT06A}
\Xi_{\epsilon}\psi_{0}^{( \epsilon )}(x)=\epsilon \psi_{0}^{(\epsilon )}(x),
\end{equation}
and an excited state $\psi_{1}^{( \epsilon)}(x)$ with corresponding energy eigenvalue $E_{1}=E_{0}^{\prime}=-1$, \emph{i. e.}
\begin{equation}\label{DBT07A}
\Xi_{\epsilon}\psi_{1}^{( \epsilon)}(x)=-\psi_{1}^{( \epsilon)}(x).
 \end{equation}
An explicit, yet not particularly illuminating, expression for $\psi_{1}^{( \epsilon)}(x)$ can be found from the relation 
\begin{equation}
\psi_{1}^{(\epsilon)}(x)\propto A_{\epsilon}\phi_{0}(x),
\end{equation}
where 
\begin{equation}\label{DBT08A}
\phi_{0}(x)=\sqrt{\frac{1}{2}}\textrm{sech}(kx),
\end{equation}
is the normalized eigensolution of operator $\eta$ for its ground level $E_{0}^{\prime}=-1$, \emph{i .e.}
\begin{equation}\label{DBT09}
\eta_{\epsilon}\phi_{0}(x)=-\phi_{0}(x).
\end{equation}
The correctness of equation (\ref{DBT06A}) can be grasped immediately by observing, from definition (\ref{DBT04}) that operator $A_{\epsilon}^{\dagger}$ annihilates $1/u_{\epsilon}$, \emph{i. e.}
\begin{equation}
A_{\epsilon}^{\dagger}\Bigg(\frac{1}{u_{\epsilon}(x)}\Bigg)=0,
\end{equation}
and then plugging $1/u_{\epsilon}(x)$ at the right of (\ref{DBT05}). In a similar way, equation (\ref{DBT09}) is proven by inserting (\ref{DBT08A}) in (\ref{DBT01}). From (\ref{DBT08A}) and (\ref{DBT09}) equation (\ref{DBT07A}) can be proven with the use of the so called intertwining relation
\begin{equation}\label{DBT10}
\Xi_{\epsilon} A_{\epsilon}=A_{\epsilon}\eta,
\end{equation}
which in turn is proven by inserting (\ref{DBT02}) and (\ref{DBT05}) in (\ref{DBT10}). 

Plugging (\ref{DBT04}) in (\ref{DBTV}) gives us, after some algebra,
\begin{equation}\label{explicitV}
V_{\epsilon}(x)= \frac{2(1+\epsilon)\big(-\epsilon+\textrm{sech}^{2}x\textrm{ sinh}^{2}\sqrt{\vert\epsilon\vert}x\big) }{\big( \textrm{tanh }x\textrm{ sinh}\sqrt{\vert \epsilon\vert}x-\sqrt{\vert \epsilon\vert}\ \textrm{cosh}\sqrt{\vert \epsilon\vert}x\big)^{2}}
\end{equation}
as an explicit expression for the $V_{\epsilon}$ potentials.
 
From (\ref{explicitV}) it is found that relations 
\begin{equation}\label{MAX}
V_{\epsilon}(x=0)=2\epsilon+2
\end{equation} 
and 
\begin{equation}\label{golden21}
\frac{d^{2}V_{\epsilon}}{dx^{2}}(x=0)=4(3+4\epsilon+\epsilon^{2}),
\end{equation} 
stands for all values of parameter $\epsilon$.
From (\ref{golden21}) we see that for values of parameter $\epsilon$ in the interval
\begin{equation}
-3<\epsilon< -1
\end{equation}
the $V_{\epsilon}$ potential is a double-well. Relation (\ref{MAX}) divides the DWP's just obtained in two different classes. First, in DWP's for which $-2<\epsilon<-1$ the ground energy level lies below the classical separatrix (Figure 1). On the other hand, if $-3<\epsilon \leq 2$ there is no energy level below the classical separatrix energy (Figure 2).  Notice that for this later type of potentials there will be no classically forbidden regions for any (stationary or non-stationary) solution, i. e. strictly speaking there will be no tunneling between wells.
\begin{figure}[!htb]
\centering
\includegraphics[scale=1.2]{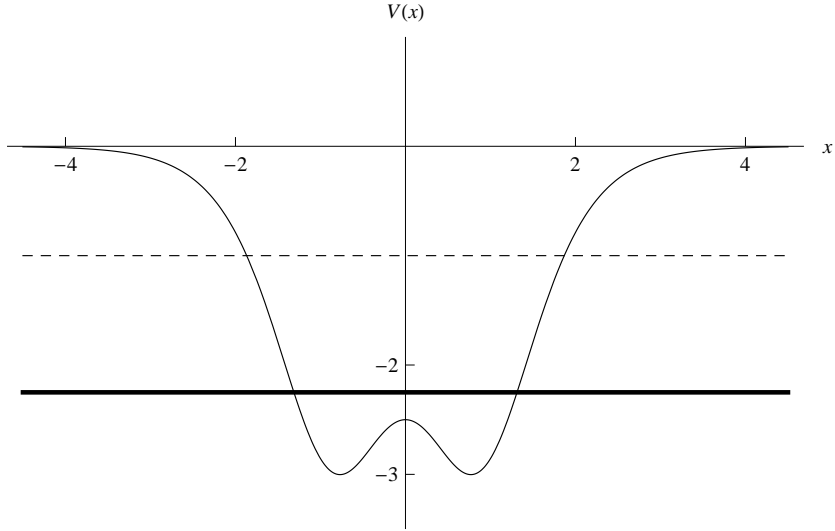}
\caption{An example of a $V_{\epsilon}$ potential, where the ground level lies above the central barrier. In this case $\epsilon=-2.25$. The first excited level $E_{1}=-1$ is shown as a dashed line, and the ground level $E_{0}=\epsilon=-2.25$ appears as a thick line.}
\label{fig:1GFH}
\end{figure}
If condition 
\begin{equation}\label{COND.L.C.}
\vert \epsilon\vert -1\ll 1
\end{equation}

is imposed, so that the lowest lying levels of $\Xi$ conform an effective two-level system, then the ground level is necessarily below the barrier, \emph{i. e.}
\begin{equation}
2\epsilon+2>\epsilon ,
\end{equation}
Also, condition (\ref{COND.L.C.}) warrants that the Leggett-Caldeira non-stationary solution
\begin{equation}
\psi_{LC}(x,t)=\frac{1}{\sqrt{2}}\Bigg(e^{-i\epsilon t}\psi_{0}^{(\epsilon )}(x)+e^{i t}\psi_{1}^{(\epsilon)}(x)\Bigg)
\end{equation}
will oscillates between wells, tunneling through the central barrier.\cite{Munoz2013} 

On the other hand, in Hamiltonians $\Xi_{\epsilon}$ with  factorization energies in the interval  $-3<\epsilon \leq 2$  the ground level  will be isolated from the rest of the spectrum by a gap wider than any gap between the rest of the levels. That is, the effective low-lying two-level approximation is inapplicable to the $\Xi_{\epsilon}$ with  factorization energies in this interval.

%
%
\section{Discussion}
The study of shallow double wells in molecular and atomic physics could be justified by the models of superconducting transitions in layered and composite superconductors mentioned in Section I. In the case of the double well model of the superionic transition of AgI, the potential changes from an asymmetric double well to a symmetric one, with the Ag migrating from the deeper well to the position of higher symmetry.\cite{Rakitin96, Kozyrev2010} This suggest to us that the symmetric double well of the this model conforms to the description of  shallow double wells given in the previous pages, as the probability density peak of the Ag then coincides with the higher symmetry position. 

Indeed, elementary considerations shows that for any analytical even double well potential (DWP from now on) $V(x)=V(-x)$ the ground state probability density $\rho_{0}(x)$ is bimodal (has two maxima) if the ground energy level is below the energy of the classical separatrix. As a matter of fact, from the Schroedinger time-independent equation one gets for an even double well    
\begin{equation}\label{Int.03}
\frac{d^{2}\rho_{0}}{dx^{2}}(x=0)=2(s-\epsilon)\rho_{0} (x=0),\quad (\hbar=2m=1)   
\end{equation}
where $s=V(x=0)$ is the separatrix energy and $\epsilon$ is the ground level energy, so that $\rho_{0}$ has  a minimum at $x=0$ if $s\geq\epsilon$. On the other hand, in the examples discussed in the present Letter, $\rho_{0}(x=0)\neq 0$, so that in this examples  the probability density has a maximum at the symmetry center of the potential whenever $\epsilon>s$. 

Yet there may exist at least one more reason for the study of shallow double wells in one-dimensional quantum mechanics, namely as a test for quantum theory in the mesoscopic realm.\cite{Leggett2002} Indeed, the existence of a non-degenerate ground state in a one-dimensional double-well potential with a central barrier of macroscopic proportions would be a prediction that would certainly clash with classical physics, as classical considerations demand for such system  the existence of two localized, parity-breaking, stable equilibrium states for a double-well. And vice versa, the experimental observation of spontaneously broken parity symmetry in, for example, an effective one-dimensional nanometric heterostructure, would be a strong indication of a failure of quantum theory in describing nanometric systems. Notice that this contradiction is ``solved" in MQC by the existence of a single non-stationary state that oscillates between two nearly degenerate stationary levels, a state of affairs that can be interpreted alternatively as the periodical transitions of the system between two degenerate states with a macroscopically different values of a measurable quantity. In an hypothetical macroscopic version of one of the shallow DWP's we have presented in this Letter no such ``amicable" solution between classical and quantum predictions would be possible.

Finally let us comment that it may be possible to construct double well potentials with properties different from those of the better known deep wells and also different from those of the shallow wells investigated in this Letter. In particular, we know of no theorem that explicitly excludes the possibility of a DWP with a two-level low-lying system above the barrier height (the classical separatrix energy level). If an example of such a potential could be presented, it would challenge the generally held notion that "there is no MQC without macroscopic tunneling." Thus, the search for such a no-go theorem (or its counterexample) is interesting at least from a purely theoretical point of view.   

\section{Conclusions and outlook}
There are at least two distinct classes of models with one-dimensional double-well potentials for a single structureless particle in non-relativistic quantum mechanics: those with bimodal ground probability density, endowed with a two-level low-lying system and associated oscillatory phenomena, and those for which the concept of tunneling results irrelevant, in which the  ground probability density is endowed with one single central peak. The shallow double well of LaO$_{0.5}$F$_{0.5}$BiS$_{2}$ explicitly belongs to this later case,\cite{Wanetal} along with the shallow potentials presented in this Letter, while examples of the former class appear, for example, in the textbook description of the ammonia molecule, and most notably in the description macroscopic quantum coherent systems. The study of shallow, and other less known types of DWP's may be prove itself to be worthwhile both for theoretical and/or applied physics.  

\section{Aknowledgements}
The support of CONACYT (Mexico) is duly acknowledged by all authors. ESH also acknowledges the support of COFFA -IPN and EDI-IPN. RMV gratefully acknowledges the financial support of FICSAC (UIA) and the sabbatical leave program of UACM, as well as the warm hospitality of \emph{Departamento de F\'{i}sica y Matem\'{a}ticas} of UIA.  

\end{document}